# Magneto-optical imaging and transport properties of FeSe superconducting tapes prepared by diffusion method

Q. P. Ding,[1,2] S. Mohan,[1] Y. Tsuchiya,[1] T. Taen,[1] Y. Nakajima,[1,2] and T. Tamegai[1,2]

[1] *Department of Applied Physics, The University of Tokyo, 7-3-1 Hongo, Bunkyo-ku, Tokyo 113-8656, Japan*
[2] *JST, Transformative Research-Project on Iron Pnictides (TRIP), 7-3-1 Hongo, Bunkyo-ku, Tokyo 113-8656, Japan*

High-quality FeSe tapes have been prepared by using a diffusion method. By adjusting the reaction temperature and the thickness systematically, a transport critical current density as high as 600 A/cm$^2$ at 4.2 K under self field is achieved. Magneto-optical images indicate considerable distribution of $T_c$ and weak-link features in the FeSe tape. This study gives some insight into how to improve the quality and performance of FeSe tapes and other superconducting tapes prepared through diffusion method further.

Discovery of iron-based superconductors have attracted a lot of attention due to their unique properties.[1] Through these studies, a binary superconducting compound FeSe (11) with the simplest structure has been discovered soon.[2] FeSe has several advantages compared with iron-pnictide superconductors, such as lower toxicity and less sensitivity to atmosphere of starting materials. Research activity directed toward the applications of this superconductor has been under way because of its high upper critical field.[3] Superconducting wires and tapes of iron-based superconductors have been fabricated using powder-in-tube (PIT) method.[4-8] However, there are several disadvantages of PIT method. In PIT method, rolling and drawing processes are needed to make the raw materials denser. In addition, the raw materials should be very fine and mixed thoroughly to achieve the homogeneity in chemical compositions. The reactivity between the sheath and the superconducting material is another issue. Since FeSe is very reactive with most of metals, only Fe sheath has been reported till now. For Fe sheath, either it also works as reacting material (*in-situ* method),[6-8] or FeSe$_{1+\delta}$ should be used to compensate the reaction with sheath (*ex-situ* method).[9,10] Diffusion method is used for a compound, composed of a high melting point component and a low melting point component, which forms new phases stable at the reaction temperature. It was originally adapted for the fabrication of A15 compounds, such as Nb$_3$Sn and V$_3$Ga.[11] It only required a shorter reaction time than conventional sintering processes, and was attempted for the fabrication of high-$T_c$ superconducting cuprates and MgB$_2$ wires and tapes.[12,13] Recently, fabrications of FeSe wires and tapes by diffusion method have also been reported.[6,14] Here we present the fabrication of high-quality FeSe tapes through diffusion method. We tried to improve the performance of FeSe tapes by optimizing a set of parameters such as the reaction temperature and the thickness systematically.

We used Se grains and 0.3 mm thick iron sheets as starting materials. Iron sheets were cut into tapes with dimensions ~ 50×5 mm$^2$. Proper amounts of Se grains and iron tapes were sealed in an evacuated quartz tube. The sealed tapes and grains were put into a muffle furnace and heated up to temperatures between 600$^{\circ}$C and 800$^{\circ}$C with a ramping rate of 100 $^{\circ}$C/h, and kept at this temperature for 12 h. Then the furnace was switched off and cooled to room temperature naturally. The crystal structures, bulk magnetization, microstructures of the samples were characterized by using X-ray diffractometer, superconducting quantum interference device SQUID magnetometer, and Scanning Electron Microscope (SEM), respectively. Resistivity and current-voltage (*I-V*) measurements were performed by four-probe method with silver paste for contacts. *I-V* measurements were performed by immersing the tape into liquid helium. For local magnetic characterization, magneto-optical (MO) imaging was applied. A Bi-substituted iron-garnet indicator film is placed in direct contact with the sample, and the whole assembly is attached to the cold finger of a He-flow cryostat and cooled down to 5 K. MO images are acquired by using a cooled CCD camera with 12-bit resolution. To enhance the visibility of the local magnetic induction, a differential imaging technique is employed.[15,16]

Fig. 1(a) shows the XRD patterns of as-prepared tapes. For tapes prepared at 600$^{\circ}$C, there always appears a hexagonal phase. For tapes prepared at 800$^{\circ}$C, the hexagonal phase only appeared in the tapes with a thickness more than 125 μm. For tapes with a thickness less than 50 μm prepared at 800$^{\circ}$C, all the peaks are well indexed using a space group of P4/nmm. The compounds crystallize in a tetragonal structured FeSe. We also attempted to prepare Fe(Te,Se) tapes by using the same method. However, the Fe(Te,Se) tape is not superconducting. The XRD pattern of Fe(Te,Se) tape is also shown in Fig. 1(a), which is not consistent with that of the Fe(Te,Se) polycrystals.[17] Although there are no impurity phases in the FeSe tape with a thickness of 25 μm prepared at 800$^{\circ}$C, optical image (Fig. 1 (b)) indicates the surface of the tape is very rough. These wavy structures will degrade the performances of the tape. Shown in Fig. 1(c) is an optical image of the tape with a thickness of 50 μm prepared at 800$^{\circ}$C. The surface is much smoother than the tape in Fig. 1(b). Below all characterizations and measurements were performed on the tape with a thickness of 50 μm prepared at 800$^{\circ}$C.

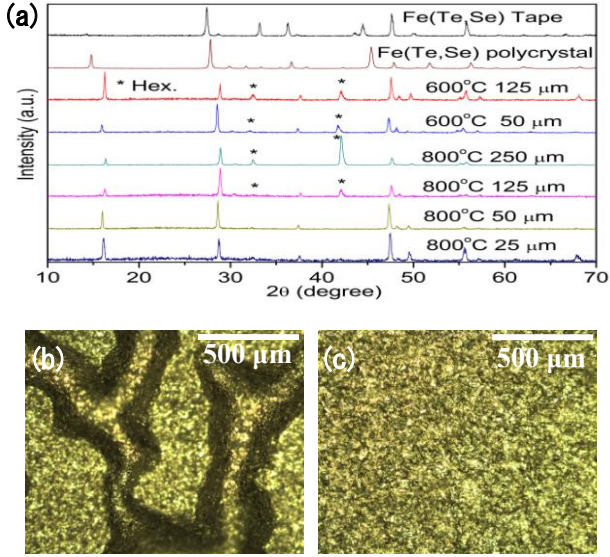

**Fig.1.** (a) XRD patterns of FeSe tapes and Fe(Te, Se) tape with different thicknesses prepared at different temperatures. (b) and (c), photographs of FeSe tapes with thickness 25 μm and 50 μm prepared at 800°C, respectively.

Magnetic hysteresis curves of the FeSe tape are shown in Fig. 2(a). Inset of Fig. 2(a) shows temperature dependences of zero-field-cooled (ZFC) and field-cooled (FC) magnetization at 5 Oe of the tape. The tape shows an onset of diamagnetism at around 8 K. By using the Bean model with an assumption of field–independent $J_c$, intragranular critical current density $J_c^{intra}$ for polycrystalline samples can be evaluated from the magnetization hysteresis loops. According to the Bean model, $J_c^{intra}$[A/cm$^2$] is given by $J_c^{intra} = 30 \Delta M/d$, with an assumption that intergranular critical current is zero, where $\Delta M$[emu/cc] is $M_{down} - M_{up}$, $M_{up}$ and $M_{down}$ are the magnetization when sweeping field up and down, respectively, $d$[cm] is the average diameter of the grains in the polycrystalline sample.[18] Fig. 2(b) is an SEM image of this tape. From this image, a typical grain size is ~ 20 μm or less, and there are many cracks between grains. $J_c^{intra}$ estimated from M-H curve is ~ $1.0 \times 10^5$ A/cm$^2$ at 5 K under zero field.

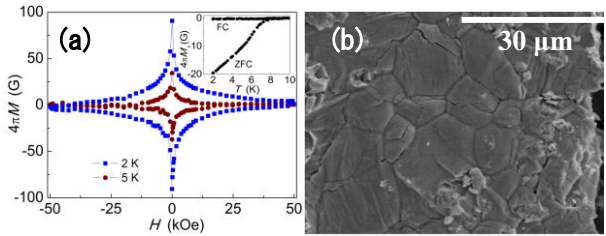

**Fig. 2.** (a) Magnetic field dependence of magnetization of FeSe tape at 2 K and 5 K. Inset: Field-cooled (FC) and zero-field-cooled (ZFC) of magnetization measured at 5 Oe. (b) SEM image of the FeSe tape.

Fig. 3(a) is an SEM image of a part of this FeSe tape physically separated from the iron core. The sample dimensions are $875 \times 775 \times 50$ μm$^3$. Figs. 3(b)-3(e) depict MO images of the FeSe tape in the remanent state after applying a 500 Oe field for 0.25 seconds which was subsequently reduced down to zero at temperatures between 5 K and 8 K. In these figures, the bright regions correspond to the trapped flux in the sample. These images are similar to the MO images of 1111 and low-temperature synthesized FeTe$_{0.5}$Se$_{0.5}$ polycrytals.[17,19-21] At all temperatures, bright intensities are restricted in small regions, implying that the intergranular current density is much smaller compared with the intragranular current density. The intragranular current density decreases gradually as the temperature is increased towards $T_c$. The MO image at 8 K clearly shows that there are considerable distributions of $T_c$ in the sample. Fig. 3(f) shows the magnetic induction profiles along the dashed line in Fig. 3(b). From the magnetic induction profile we calculated the intragranular critical current density. In this calculation, we roughly estimate the intragranular critical current density by $J_c^{intra} \sim dB/dx$. The estimated $J_c^{intra}$ is ~ $1.1 \times 10^4$ A/cm$^2$ at 5 K for typical grains. Compared with $J_c^{intra}$ calculated from the M-H curve, the value estimated from MO image is much smaller. The reason for this difference may be similar to the case of the FeTe$_{0.5}$Se$_{0.5}$ polycrystals, where grains far from the surface cancel part of magnetic induction.[17]

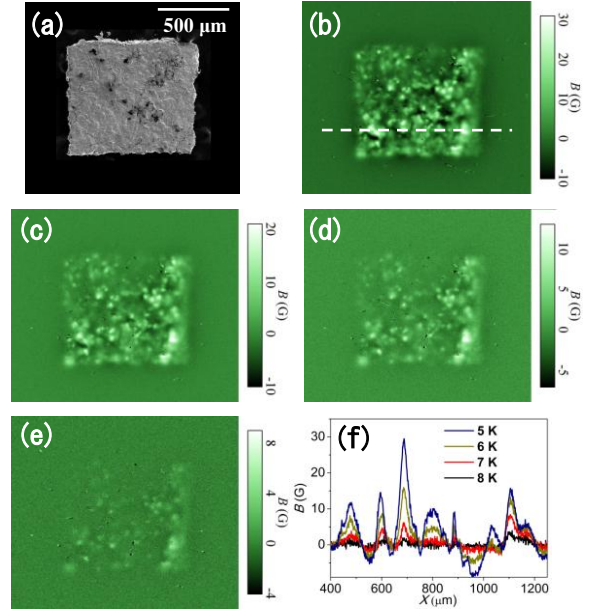

**Fig. 3.** (a) SEM image of the FeSe tape used for MO imaging. Differential MO images in the remanent state of the FeSe tape at (b) 5 K, (c) 6 K , (d) 7 K, and (e) 8 K, after cycling the field up to 500 Oe for 0.25 seconds. (f) The local magnetic induction profiles at different temperatures taken along the dashed lines in (b).

Resistive transitions under magnetic fields of $H = 0$, 10, 20, 30, 40 and 50 kOe are shown in Fig. 4(a). The transition shifts to lower temperatures accompanied by a slight increase in the transition width when the field is increased. Inset of Fig. 4(a) shows the variation of the upper critical field $H_{c2}$ with reduced temperature $t = T/T_c^{onset}$ for the FeSe tape. The values of $H_{c2}$ were defined as the field at midpoints of the resistive transitions. The slope of $H_{c2}$ at $T_c$ is -26.4 kOe K$^{-1}$. Using the Werthamer–Helfand–Hohenberg formula,[22] $H_{c2}(0) = -0.69T_c|dH_{c2}/dT|_{T=Tc}$, the value of $H_{c2}$ at $T = 0$ K is 243 kOe. This value is similar to that of single crystals and other reported FeSe wire and tapes,[3,6,14] and much larger than that of the first report of FeSe.[2] We have used the Ginzburg–Landau formula to extract the superconducting coherence length ($\xi$), $\xi = (\Phi_0/2\pi H_{c2})^{1/2}$, where $\Phi_0 = 2.07 \times 10^{-7}$ G cm$^2$. At zero temperature the coherence length $\xi$ is calculated as 3.64 nm.

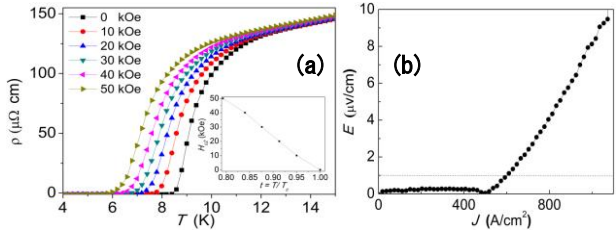

**Fig. 4.** (a) Field dependence of the resistivity of FeSe tape around $T_c$. Inset shows the upper critical field $H_{c2}$ versus normalized temperature determined by the midpoint of the resistive transition. (b) E–J characteristics of FeSe tape at 4.2 K.

We also investigated the transport $J_c$ of this FeSe tape. The distance between two voltage contacts is 820 μm. The E–J characteristic at 4.2 K under self field is shown in Fig. 4(b). Current density was obtained by dividing the current by the cross sectional area of FeSe tape. To define transport $J_c$, we adopt $E = 1$ μV/cm as a criterion for the E–J curve here. Transport $J_c$ as high as 600 A/cm$^2$ is achieved at 4.2 K in this FeSe tape. This value is 5 times as that of reported FeSe tape,[14] and also twice as that of the best single-core FeSe wire.[6]

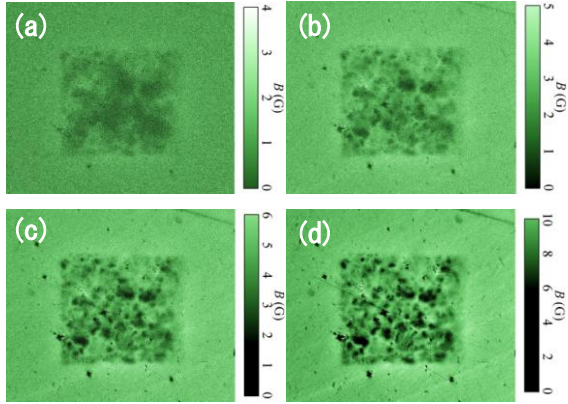

**Fig. 5.** MO images of flux penetration into FeSe tape at (a) 1 Oe, (b) 3 Oe, (c) 5 Oe, and (d) 10 Oe, after zero-field cooling down to 5 K.

Figs. 5(a) to 5(d) reveal the penetration of vortices at 5 K in the FeSe tape. In Fig. 5(a), at low field of 1 Oe, most parts of the sample are still in the Meissner state. Magnetic flux penetrates intergranular regions when the field is increased to 3 Oe (Fig. 5(b)). When the field is increased further (Figs. 5(c) and 5(d)), the shapes of the superconducting grains are more visible. These images demonstrate the effect of weak links between superconducting grains even at low fields.

In superconducting wires and tapes, both $J_c$ and critical current ($I_c$) are important parameters for applications. Using our diffusion method, it is easy to increase $I_c$ by simply increasing the width of the starting Fe tape. By using multifilamentary structures, $J_c$ can be further increased to several times as single-core structure.[6] The quality and performance of FeSe tapes can also be improved by further optimizing the ramping rate to control the vapor pressure, and the cooling process. Introducing an annealing process may reduce the distribution of $T_c$ thus makes the FeSe tape more homogenous. Use of structured Fe tapes may promote the texturing of final FeSe tapes.

To summarize, we have reported X-ray diffraction, magnetization, resistivity, transport critical current density, and magneto-optical measurements of high quality FeSe tapes fabricated by diffusion method. A transport $J_c$ as high as 600 A/cm$^2$ at 4.2 K under self field is achieved. This value is 5 times as that of reported FeSe tape, and also twice as that of the best single-core FeSe wire. Howerver it is more than two orders less than the intragranular critical current density. The $H_{c2}$ at $T = 0$ K is estimated as 243 kOe, which is similar to that of single crystals. MO images indicate considerable distribution of $T_c$ and weak-link features in the FeSe tape. Reduction of cracks between superconducting grains will be one of the most effective ways to improve the performance of superconducting tapes prepared by diffusion method.